\documentclass{elsart}

\usepackage{float}
\usepackage{color}
\usepackage{subfigure}
\usepackage[pdftex]{graphicx}
\IfFileExists{url.sty}{\usepackage{url}}
                      {\newcommand{\url}{\texttt}}

\newcommand{\pythia}{{\textsc{Pythia }}}

\newcommand{\filetext}[1]{\texttt{#1}}
\newcommand{\subroutinetext}[1]{\texttt{#1}}
\newcommand{\shellcommand}[1]{\texttt{#1}}

\newenvironment{entry}%
{\begin{list}{}{\setlength{\topsep}{0mm} \setlength{\itemsep}{0mm}
\setlength{\parskip}{0mm} \setlength{\parsep}{0mm}
\setlength{\leftmargin}{20mm} \setlength{\rightmargin}{0mm}
\setlength{\labelwidth}{18mm} \setlength{\labelsep}{2mm}}}%
{\end{list}}

\newcommand{\iteme}[1]{\item[\texttt{#1}\hfill]}

\begin{document}
\begin{frontmatter}
\title{EtabFDC:\\ an $\eta_b$ event generator in hadroproduction at LHC\\}
\author[gucas,tpcsf]{Cong-Feng Qiao},
\author[gucas]{Jian Wang},
\author[ihep]{Jian Xiong Wang},
\author[gucas]{Yangheng Zheng}
\address[gucas]{Graduate University of Chinese Academy of Sciences, Beijing, 100049, China}
\address[ihep]{Institute of High Energy Physics, CAS, Beijing 100049, China}
\address[tpcsf]{Theoretical Physics Center for Science Facilities, CAS, Beijing 100049, China}

\begin{abstract}

The EtabFDC is a matrix-element event generator package for
$\eta_b$ hadroproduction at LHC. It generates $pp\rightarrow\eta_b X$
events for all possible parton-level $2\rightarrow2$ leading-order
processes with three commonly used $\eta_b$ decay channels being
implemented. The \pythia interface is used for parton-shower and
hadronization to obtain the hadronic events. The FORTRAN codes of
this package are generated by FDC (Feynman Diagram Calculation)
system automatically.

\end{abstract}
\end{frontmatter}
\noindent
\textbf{Program summary}\\

\emph{Program Title:} EtabFDC (Version 1.01)



\emph{Operating system:} Linux (with GNU FORTRAN 77 compiler)

\emph{Programming language:} FORTRAN 77

\emph{External libraries:} CERNLIB 2001 (or CERNLIB 2003)

\emph{Distribution format:} tar gzip file

\emph{Size of the compressed distribution file:} 17.1M Bytes.

\emph{Classification:} 11.1

\emph{Nature of physical problem:} This package generates
$pp\rightarrow\eta_b X$ events for partonic $2\rightarrow2$
tree-level processes with implementing three $\eta_b$ decay
channels. It then interfaces with \pythia for parton-shower and
hadronization to obtain the hadronic events.

\emph{Method of solution:} The FORTRAN codes of this package are
generated by FDC system\cite{Wang:2004du} automatically.

\emph{Typical running time:} To generate 1,000 events on a 3.2G
Intel-P4 CPU platform (with hyper-technology), it will take
approximate 20, 30 and 300 seconds for decay channel $\eta_b \rightarrow J/\psi
\gamma$, $\eta_b \rightarrow J/\psi J/\psi$ and $\eta_b \rightarrow
J/\psi c \bar{c}$ respectively.

\section{Introduction}\label{secINTRO}
As the $b\bar{b}$ ground state, $\eta_b$ has been a focus in high
energy physics since it had not been found experimentally for more
than thirty years. Many theoretical
models\cite{Bali,Ackleh,Godfrey,Marcantonio,Bali2,Brambilla,Narison,Eichten,Ebert}
estimated the properties of $\eta_b$. The commissioning LHC
experiments will provide a excellent opportunity to search for the
signal of $\eta_b$. Thus, an event generator software, that
simulates the $\eta_b$ production and decays, is necessary to
understand the reconstructed signal and backgrounds in the $pp$
collision environment. In this work, we just present such a
package, named EtabFDC, that can generate $pp\rightarrow\eta_b X$
events from leading-order (LO) processes (Fig. \ref{figANN} and Fig.
\ref{figFRA}) at parton-level. Here, the total cross-section of the
$\eta_b$ production is calculated to be $9.3\times10^{7}$Pb. Three
$\eta_b$ decay modes, that may be reconstructed by the LHC
experiments, are included in this package. They are: $\eta_b
\rightarrow J/\psi \gamma$ \cite{haogang1}, $\eta_b \rightarrow
J/\psi J/\psi$ and $\eta_b \rightarrow J/\psi c \bar{c}$
\cite{haogang2}. If the branching-ratio is provided, we can estimate
the number of produced $\eta_b$ events of a specific decay channel
in LHC for a given luminosity. For example, the theoretical
prediction of the branching-ratio of channel $\eta_b \rightarrow
J/\psi J/\psi$ is 5$\times10^{-8}$. And considering the $1 fb^{-1}$
data collected by the LHC detector in the first year, the number of
$\eta_b\rightarrow J/\psi J/\psi$ events produced is about 4650.
Under current software framework, more decay channels can be added
as the independent modules by the future demand.

Feynman Diagram Calculation package (FDC) \cite{Wang:2004du,jp2000}
is one of the projects which aimed at calculationg Feynman diagram
automatically. It can conveniently implement physical models from
the first principle, then construct their
\texttt{\emph{Lagrangian}}s, deduce the Feynman rules and generate
FORTRAN codes for calculations of physical quantities such as decay
width, cross-section and matrix-element. EtabFDC is a FDC
application and its FORTRAN codes for the calculation of the
$\eta_b$ production processes are generated by FDC automatically.
And the three decay modes of $\eta_b$ are added by manual wrok.  In
this way, many careless errors can be systematically checked and
debugged. EtabFDC has interface with \pythia to generating the
hadronic events.

\begin{figure}[H]
\centering
\includegraphics[scale=0.5]{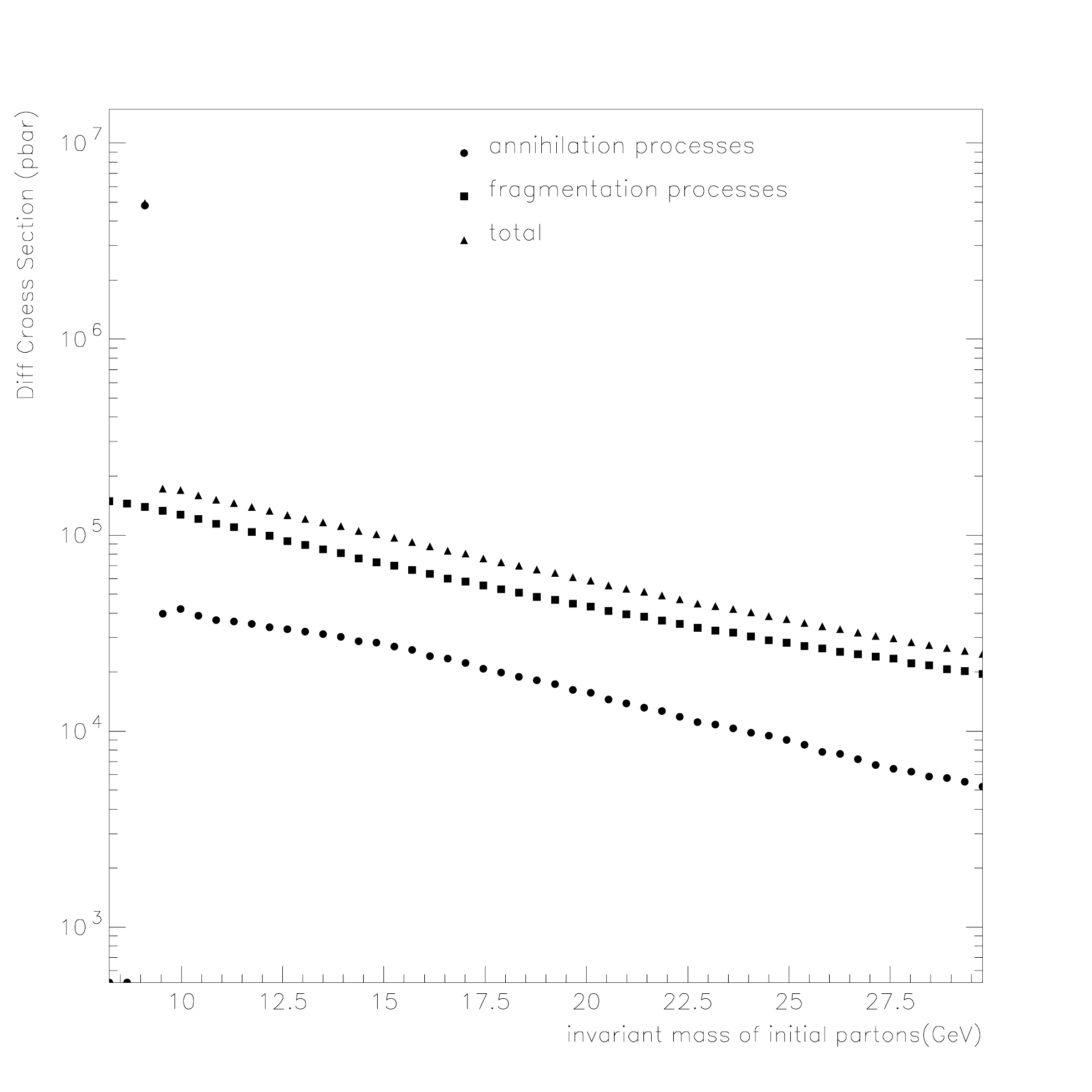}
\caption{Differential cross-section of $\eta_{b}$ production vs c.m. energy of the 2
initial partons for the $2\rightarrow2$ tree-level
processes}\label{figmas}
\end{figure}

\begin{figure}[H]
\centering
\includegraphics[scale=0.5]{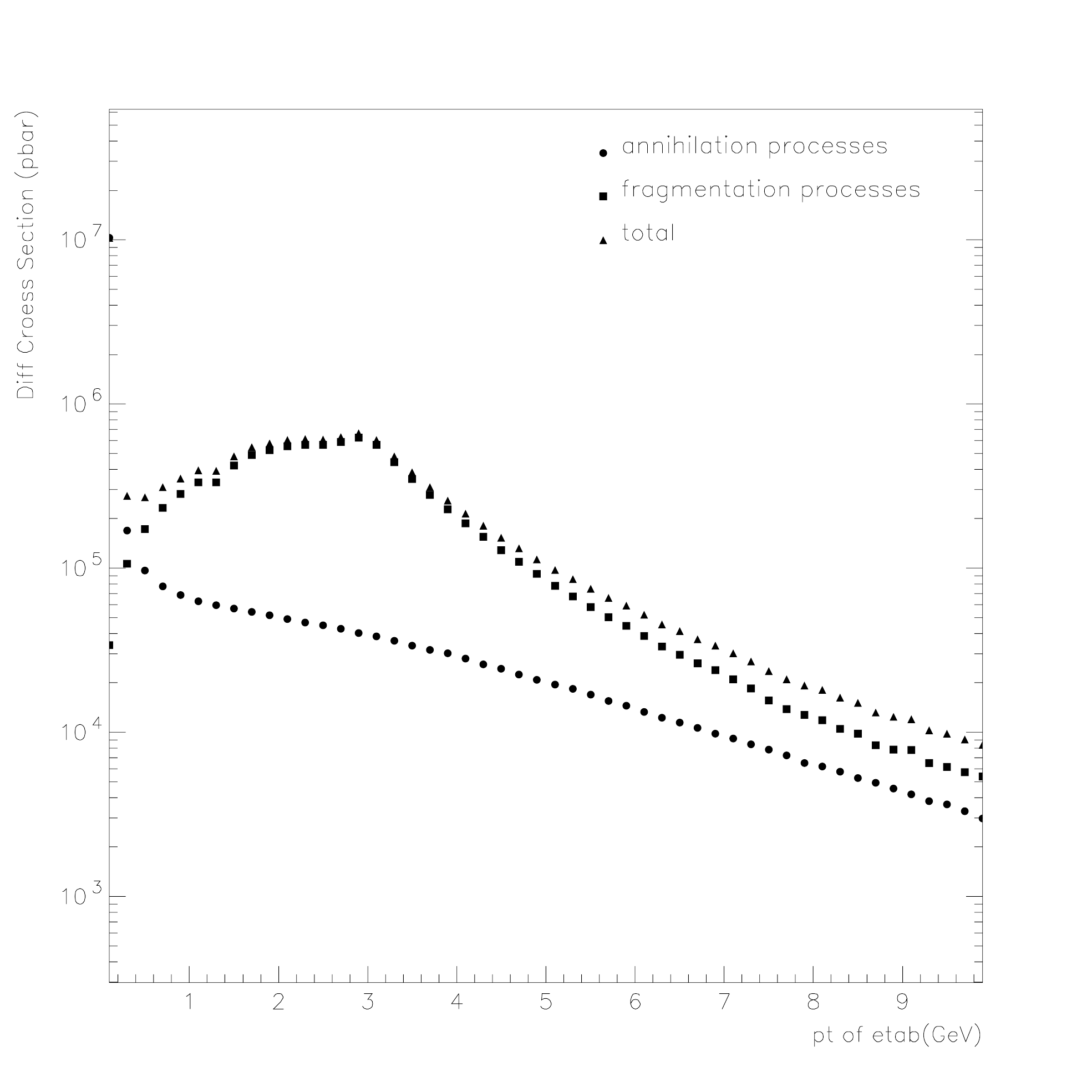}
\caption{$\eta_{b}$ production cross-section vs transverse momentum
$p_{t}$}
\label{figvpt}
\end{figure}

As a test, we plot the 2-initial-partons' centre of mass (c.m.) energy (Fig.
\ref{figmas}) and the transverse-momentum $p_t$ (Fig. \ref{figvpt})
distributions with generating 20-million $\eta_b$ events. Here, we
put the $pp$ collisions energy at $\sqrt{s}=14$ TeV. And the
$\eta_b$ mass value is set to be 9.3 GeV. In Fig. \ref{figmas}, the
peak at $9.3$ GeV corresponds to the $g+g\rightarrow\eta_{b}$
process, where the c.m. energy of the two initial gluons has to be
equal to the mass of $\eta_b$. Please also note that, in Fig.
\ref{figvpt}, a bump around 3 GeV is produced for the fragmentation
processes. The cross-section shortfall below 3 GeV is caused by a
transverse-momentum cut, $p_t
>3$ GeV, on the fragmented parton $b$ or $g$, which then form the
$\eta_b$. We obtained the expected results from the simulation
output.

This paper is organized as follows: The installation procedures and
the directory structures are described in Section~\ref{secINST}. In
Section~\ref{secImp}, we describe the details of the implementation
of EtabFDC package. Section~\ref{secHOWTOUSE} presents the usage
manual. Sections~\ref{secEVENTFILE} describes the interface and
format of the output files. The summary is presented in
Section~\ref{secCON}.

\section{Installation and the directory
structures}\label{secINST}

\subsection{Installation}\label{subsecINST}

The EtabFDC runs under Linux system and \texttt{GNU FORTRAN 77}
compiler is required (other FORTRAN compilers are not tested). The
necessary \pythia library files are included in the package for
convenience. The CERN program Library 2001 or 2003 is also required,
but not included with this package to avoid possible platform and
compiler incompatibility. The package is distributed in a compressed
file named \filetext{etabfdc.taz} and can be downloaded from FDC
homepage:

\begin{quote}
\url{http://www.ihep.ac.cn/lunwen/wjx/public_html/genpack/genpack.html}
\end{quote}

The installation steps are given below and for convenience we assume
that \footnote{Users can substitute the paths with their own working
directory while installing the EtabFDC package.}:
\begin{itemize}
\item[1.] the current user account is \texttt{user};
\item[2.] the current directory is \texttt{/home/user};
\item[3.] it is assumed that the user use \shellcommand{"csh"} ; if not , the user
should execute \shellcommand{"chsh useraccount" or "ypchsh
useraccount"} to change the shell class to cshell ;
\item[4.] CERN library should be CERN2001.
\end{itemize}

For the installation, user first need to decompress the downloaded
gzip file by typing the following command.\\
   \hspace*{3em}\shellcommand{"tar -xzvf etabfdc.taz"} \\
then home directory "etabfdc" is generated.
Secondly, get into it by "cd etabfdc" and set the correct path of CERN library in env.csh,then execute: \\
   \hspace*{3em}\shellcommand{ "source env.csh" and "installlib"}. \\
It will do the installation.

\subsection{Directory structures}\label{subsecDIR}
\begin{figure}[h]
\centering
\includegraphics[scale=0.5]{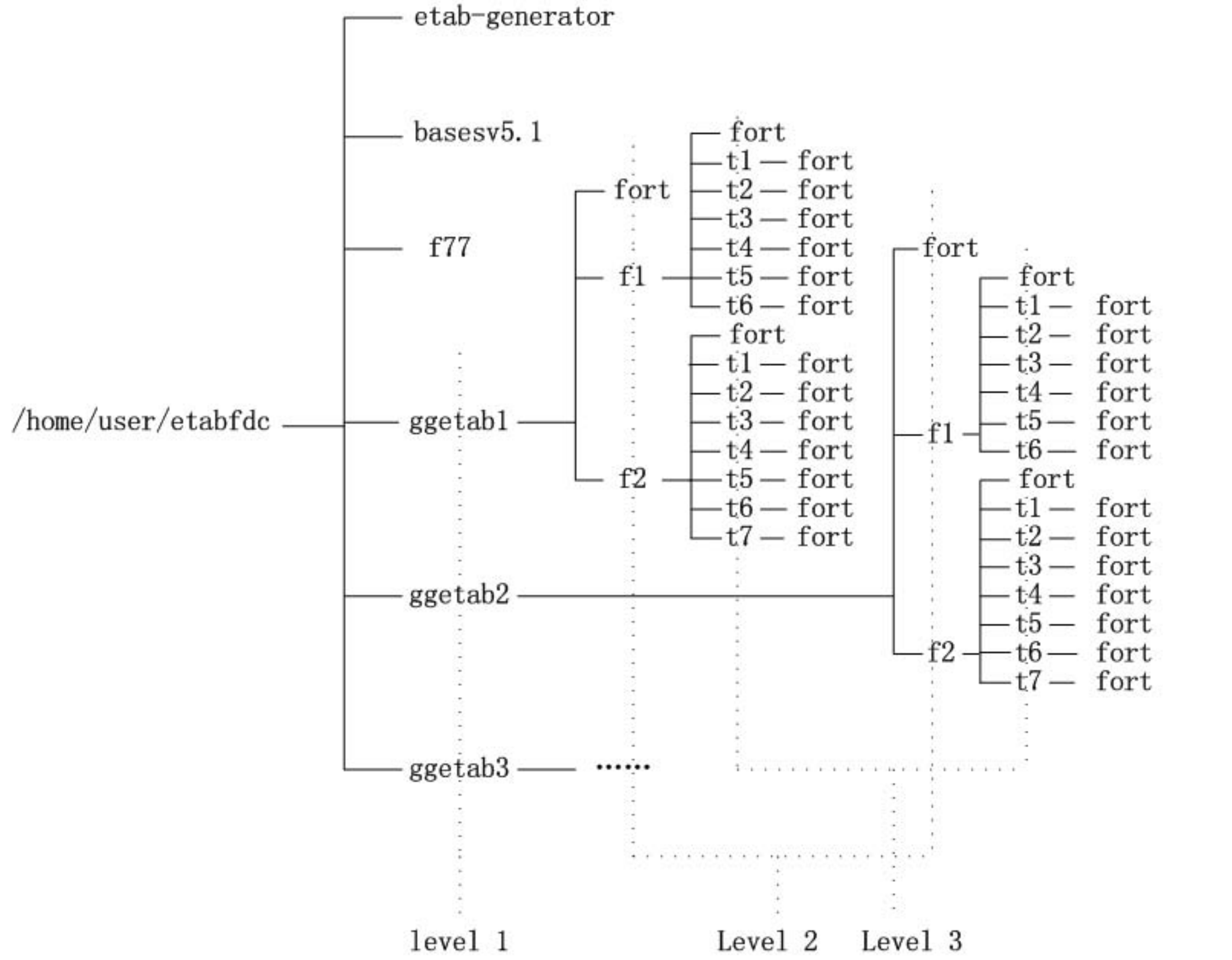}
\caption{Directory structure of EtabFDC}\label{figDIR}
\end{figure}

After the successful decompression and setting up the environment,
user can find six sub-directories (see Fig. \ref{figDIR}) in the
installation directory. They are \filetext{basesv5.1},
\filetext{etab-generator}, \filetext{f77}, \filetext{ggetab1},
\filetext{ggetab2} and \filetext{ggetab3}. The directory
\filetext{basev5.1} stores \texttt{BASES} libraries used for
Monte-Carlo calculation of the cross-section integral. The directory
\filetext{f77} are the common-shared computational tools for
$\eta_b$ decay processes and the directory \filetext{etab-generator}
contains the $\eta_b$ production information.

\filetext{ggetab1},\filetext{ggetab2} and \filetext{ggetab3} are
three important directories that contain the parton-level
annihilation and fragmentation channels, and correspond to three
different $\eta_b$ decay channels: $\eta_b \rightarrow J/\psi
\gamma$, $\eta_b \rightarrow J/\psi J/\psi$ and $\eta_b \rightarrow
J/\psi c \bar{c}$, respectively. All the compiled executable
programs and the input files for each decay channel are stored here.
Fully installed package substructure are shown in Fig. \ref{figDIR},
based on which the EtabFDC package is constructed into three running
levels.

Many generators (such as \pythia) are programmed in such a way that
all implemented processes are packed into a single FORTRAN file and
some variables are used to enable or disable these processes.
Nevertheless, EtabFDC works in a different way. There are three
running levels for generating $pp\rightarrow\eta_b X$ events. And
each running level consists of a series of stand-alone executable
programs and input-files which are stored in different
sub-directories. The introduction to the running levels and the
details of substructure of the directories will be given in the
following section.

\section{Implementation}\label{secImp}

In EtabFDC, the cross section of each LO
process is calculated by using \texttt{BASES}\cite{bases} package) at first
and these cross sections are used as the weight for each processes.
The spring in \texttt{BASES} is used to generate parton-level $\eta_b$ events for each processes.
and then the parton-level $\eta_b$ events are passed to \pythia
\cite{pythia} by Les Houches Interface Standard\cite{les} for
partonic showering and hadronization. The interface with \pythia is
shown in Fig. \ref{figPYTHIA}, and will be described in section
\ref{subsecINTERFACE}.

We implement the EtabFDC in such a flexible way that users have
options to select on generating events from three running levels.

\subsection{Generality about running levels}\label{subsecWHAT}

As shown in Fig. \ref{figDIR}, There is a set of stand-alone
programs to generate events at each running level as, for all
processes at running level 1, for a certain class, annihilation or
fragmentation, of processes at running level 2, or for a specific
process at running level 3.

For the three $\eta_b$ decay channels, the programs for ``level 1"
are stored in \filetext{ggetab1/fort} , \filetext{ggetab2/fort} and
\filetext{ggetab3/fort} respectively.  The directories \filetext{f1}
and \filetext{f2} under \filetext{ggetab1}(or
\filetext{ggetab2},\filetext{ggetab3}) are called
\textbf{f-directories}, which contain the stand-alone programs for
``level 2". These programs are stored in
\filetext{f\textit{n}/fort}. \filetext{f1} or \filetext{f1} contains
the annihilation or the fragmentation processes. The directories
\filetext{t1}, \filetext{t2} and etc under f-directories are called
\textbf{t-directories} for ``level 3". These programs are stored in
\filetext{t\textit{n}/fort} and can generate unweighted events for
each specific process.

\begin{figure}[H]
\centering \subfigure [$q+\bar{q}\rightarrow\eta_{b}+\gamma$]
{\includegraphics[scale=0.6]{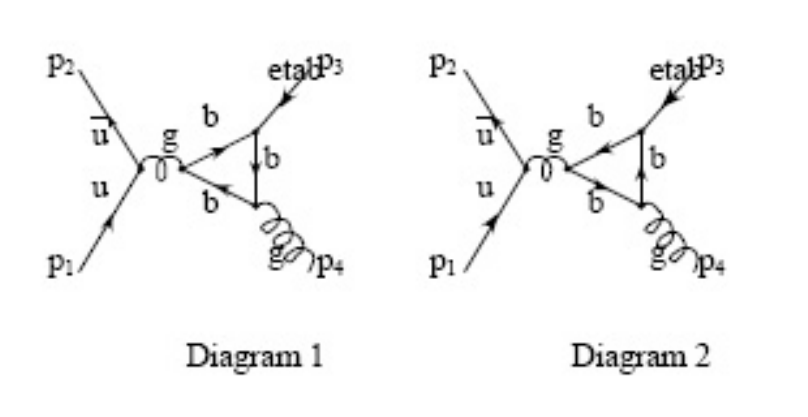}}
\subfigure[$g+q\rightarrow\eta_{b}+q$]
{\includegraphics[scale=0.6]{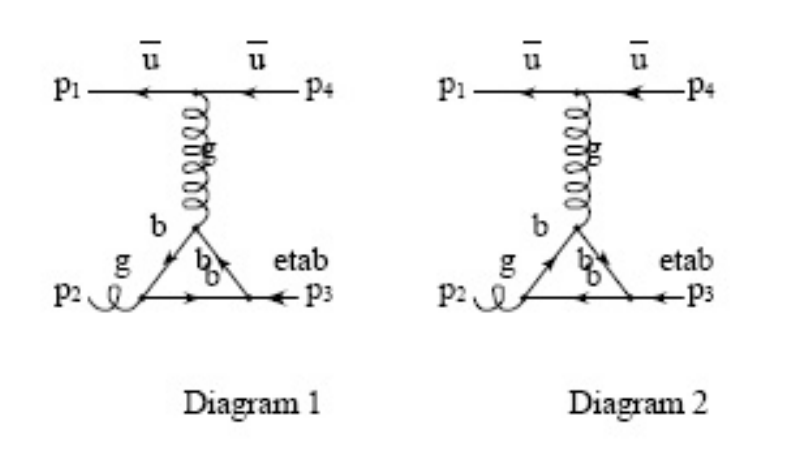}}
\subfigure[$g+g\rightarrow\eta_{b}+g$]
{\includegraphics[scale=0.55]{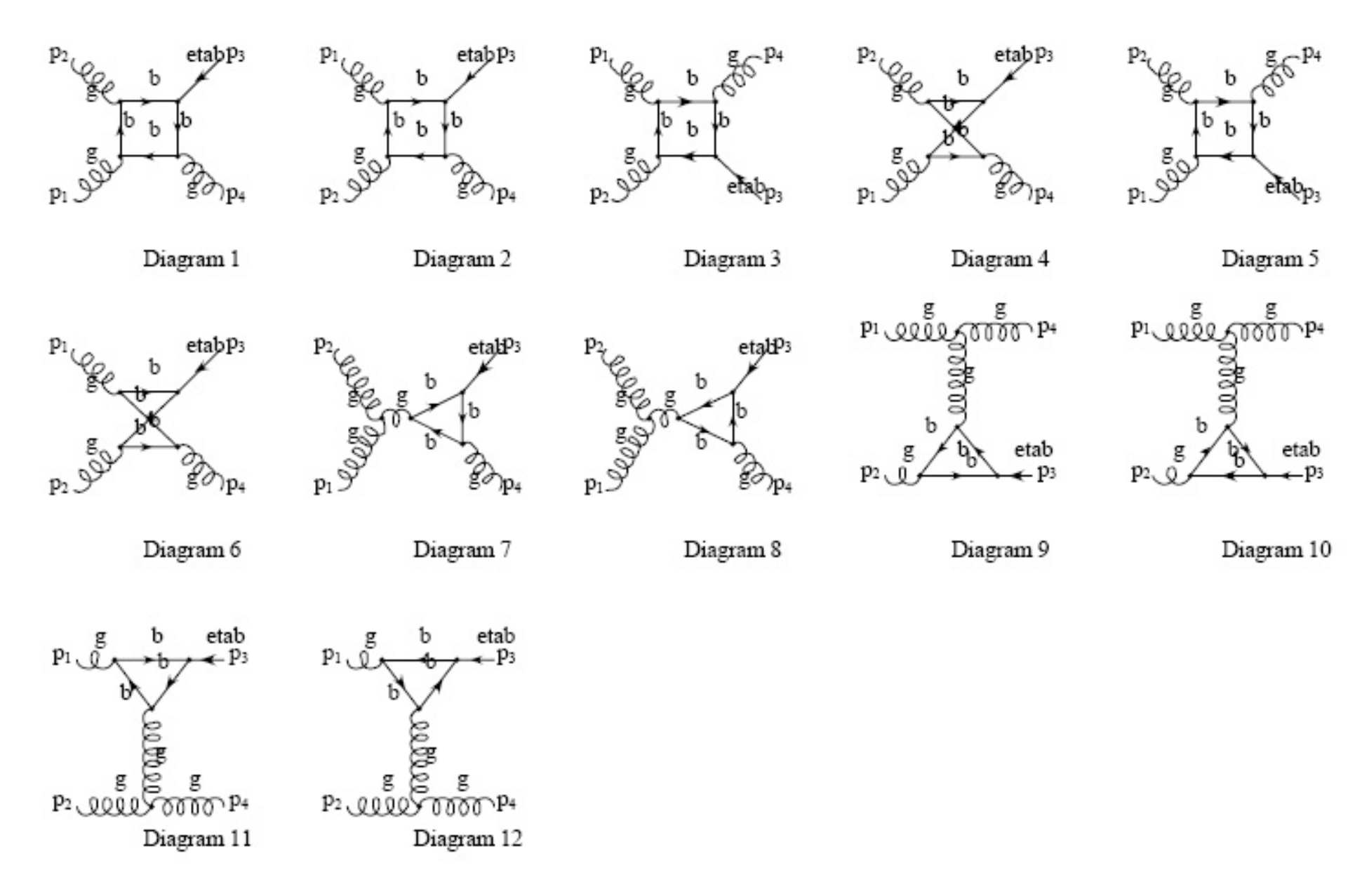}}\\
\subfigure[$g+g\rightarrow\eta_{b}$]
{\includegraphics[scale=0.3]{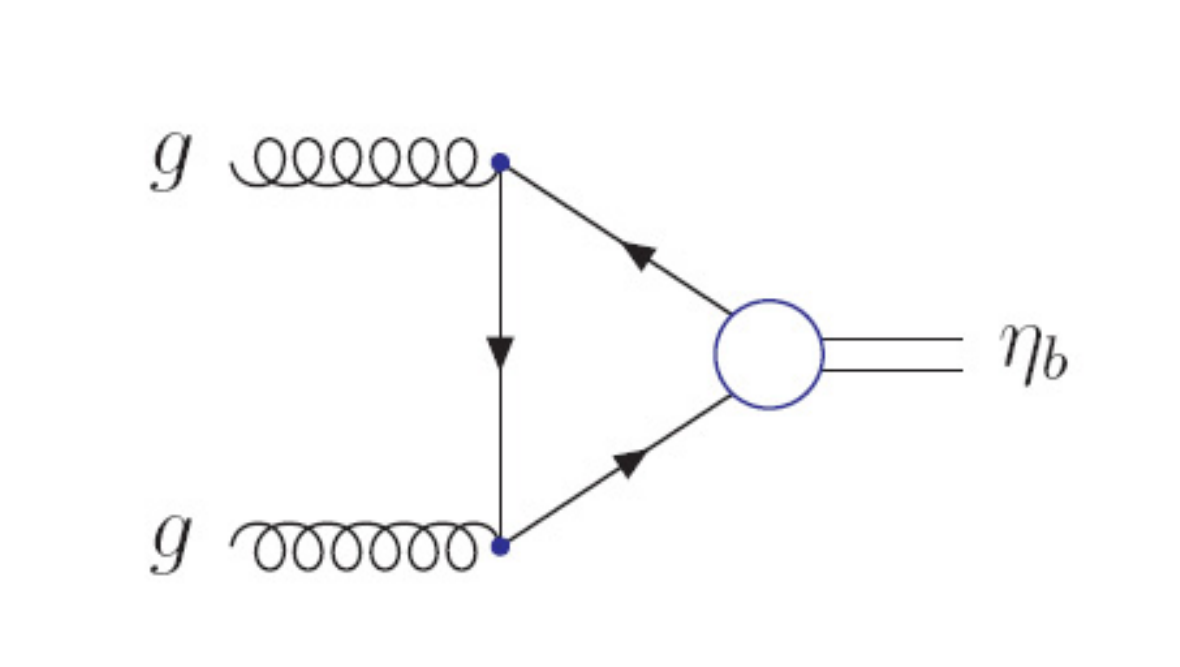}} \caption{Feynman diagrams for
the leading-order annihilation processes in the $\eta_b$
generation.}\label{figANN}
\end{figure}

In the following, the symbol ``$q$" is used to represent ``$u,d,s$" quarks.
The \textbf{t-directories} under \filetext{f1} for annihilation processes is :\\
t1 : $q+\bar{q}\rightarrow\eta_{b}+\gamma$ \\
t2 : $g+s\rightarrow\eta_{b}+s$\\
t3 : $g+d\rightarrow\eta_{b}+d$\\
t4 : $g+u\rightarrow\eta_{b}+u$\\
t5 : $g+g\rightarrow\eta_{b}+g$\\
t6 : $g+g\rightarrow\eta_{b}$.\\
The Feynman diagrams are shown in Fig. \ref{figANN} (u quark is used
as example for q here).

\begin{figure}[H]
\centering \subfigure [$q+\bar{q}\rightarrow\ g+g $]
{\includegraphics[scale=0.5]{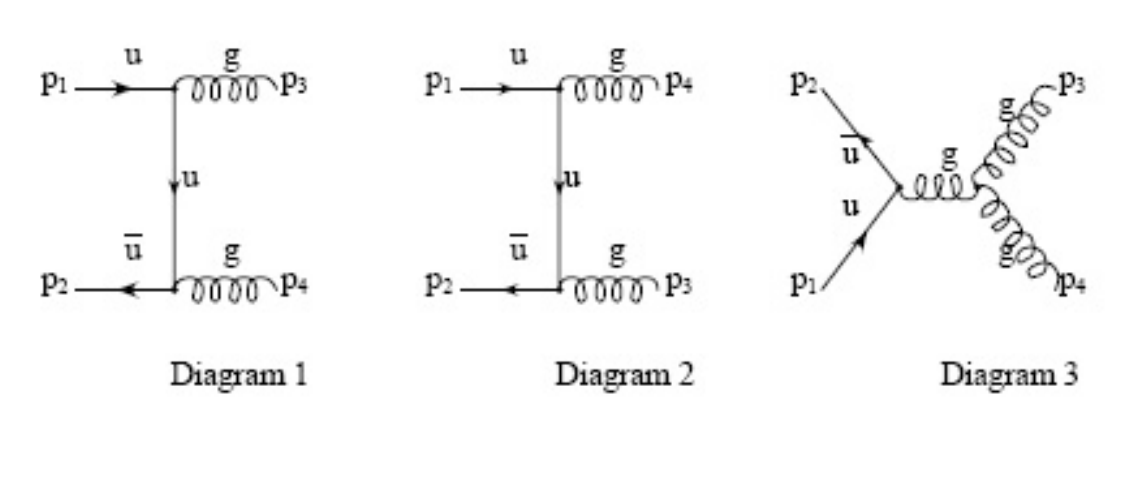}} \subfigure [$g+q\rightarrow\
g+q$] {\includegraphics[scale=0.5]{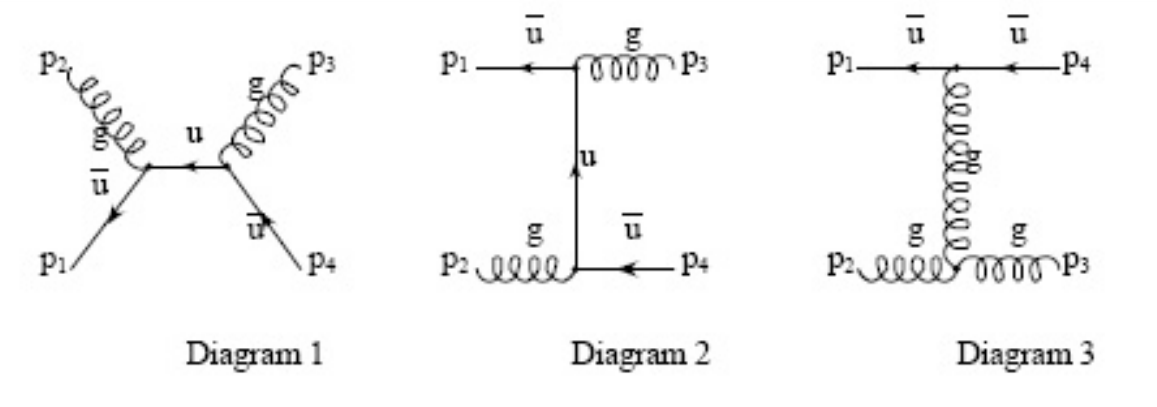}} \subfigure
[$g+g\rightarrow\ g+g$] {\includegraphics[scale=0.5]{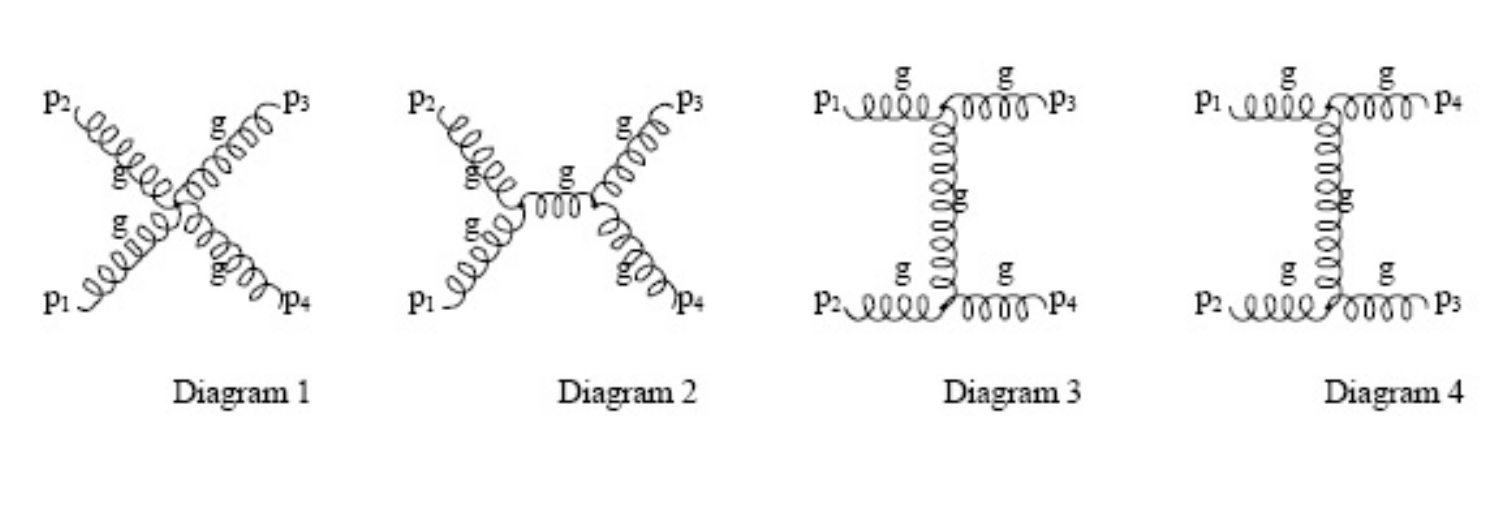}}
\subfigure [$q+\bar{q}\rightarrow\ b+\bar{b}$]
{\includegraphics[scale=0.6]{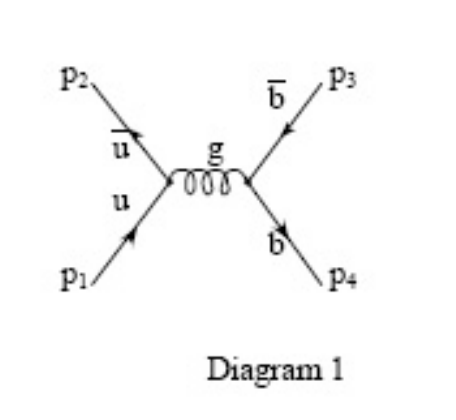}} \subfigure [$g+g\rightarrow\
b+\bar{b}$] {\includegraphics[scale=0.5]{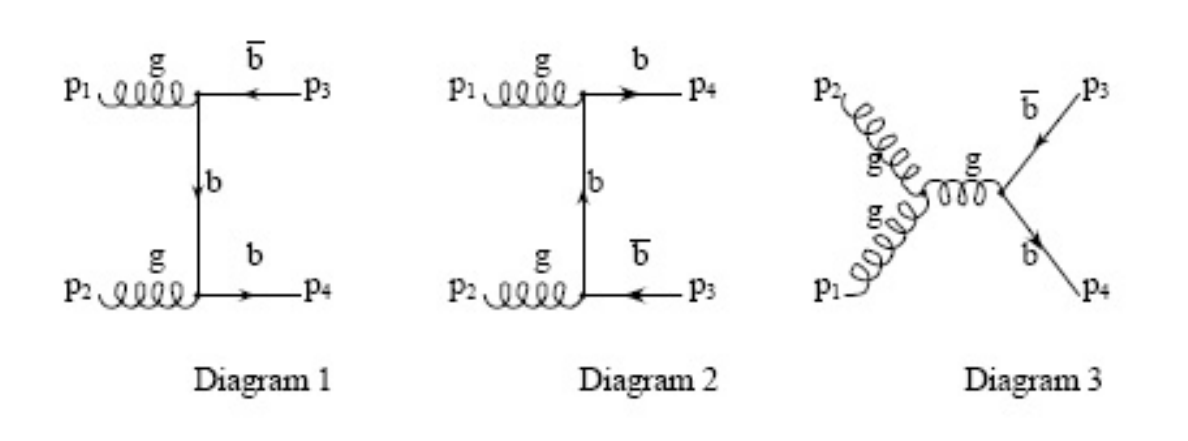}} \caption{Feynman
diagrams for the leading-order fragmentation processes in the
$\eta_b$ generation.}\label{figFRA}
\end{figure}

Similarly, the \textbf{t-directories} under \filetext{f2} for the fragmentation processes is:\\
t1 : $q+\bar{q}\rightarrow\ g+g$\\
t2 : $g+s\rightarrow\ g+s$\\
t3 : $g+d\rightarrow\ g+d$\\
t4 : $g+u\rightarrow\ g+u$\\
t5 : $g+g\rightarrow\ g+g$\\
t6 : $q+\bar{q}\rightarrow\ b+\bar{b}$\\
t7 : $g+g\rightarrow\ b+\bar{b}$,\\
Where the gluon and $b$ or $\bar{b}$ can fragment into $\eta_b$.
The Feynman diagrams are shown in Fig. \ref{figFRA} (u quark is used
as example for q here).

For each process, the value of cross-section is saved in the
file \textbf{fresult.dat} in \textbf{fort} directory under the directory of the process.

Among the three running levels, only level 3 interfaces with \pythia
directly, while level 1 and 2 interface with \pythia through level
3. Interfacing details will be discribed in section
\ref{subsecINTERFACE}.

\subsection{Running level 3}

When EtabFDC runs at level 3, it first generates unweighted
matrix-element events and then interfaces with \pythia for
parton-showering and hadronization to obtain hadronic events. In
this running level, each \textbf{t-directory} contains a specific
parton-level process. There are three executable programs and one
input-file stored in \filetext{ggetab\textit{n}/t\textit{n}/fort}.
They are listed in table \ref{tblEXE3}.
\begin{table}[h]
\caption{Executable programs in level 3} \label{tblEXE3}
\centering
  \begin{tabular}{ll}
    \hline
    \filetext{int} & calculates and stores the cross-section \\
     &  in file \filetext{fort/fresult.dat}\\

    \filetext{gevent} & generates and stores parton-level events with $\eta_b$ being decayed\\
     &  in file \filetext{fort/pdata1.dat} \\

    \filetext{fdcpythia} & interfaces with \pythia and stores the final hadronic
    events\\
    &   in file \filetext{fort/pyresult.dat} \\

    \filetext{parameter.input} & input-file in which users can specify initial
    parameters\\ &  such as running level and number of events\\
    \hline
  \end{tabular}
\end{table}

The programs \filetext{int} and \filetext{gevent} can be called by
\filetext{fdcpythia} automatically. The program \filetext{fdcpythia}
reads \filetext{parameter.input} for the initial parameters. The
format of \filetext{parameter.input} is described as: each parameter
occupies one line; in each line, the parameter name and value are
separated by a equal sign (i.e. ``='') like:
\begin{verbatim}
    parameter = value;
\end{verbatim}
the line which begins with a number sign (i.e. ``\#") is recognized
as the comment. All parameters used by EtabFDC in
\filetext{parameter.input} of running level 3 are listed
with brief explanation as:
\begin{entry}
\iteme{RUNNINGLEVEL(=1,2,3)} : To specify the running level.
When EtabFDC runs at level 3, it should be set to 3 by users.
Otherwise it will be set automatically.

\iteme{EVNTNUM} : To specify the number of the generated hadronic events.

\iteme{PARTONEVNTNUM} : The number of the generated parton-level
events. Since \pythia can reject some parton-level events, we need
to set this number greater than the final event number. Typically,
it should be set at least 20 percent more
than \texttt{EVNTNUM}.

\iteme{PRESENTEVNTNUM} : The number of generated hadronic events in the last run.
It is set automatically and should not be modified by users.

\iteme{RESET} : To control the program \filetext{fdcpythia} to run in two
different modes -- \texttt{Re-generation} mode and
\texttt{Appending} mode. In EtabFDC package, we only use the
\texttt{Re-generation} mode (\texttt{=1}), since all the generated
events are erased, and new parton-level events and hadronic events are generated.
\end{entry}

\subsection{Running level 2}

When EtabFDC runs at level 2, it generates either annihilation or
fragmentation class of parton-level events. For each class, the
cross-section of each parton-level process are used as weight of the
total generated number of events. In this level, each class of the
processes corresponds to an f-directory. There are six programs for
each class of processes which are stored in
\filetext{ggetab\textit{n}/f\textit{n}/fort} and listed in table
\ref{tblEXE2}
\begin{table}[h]
\caption{Executable programs for level 2} \label{tblEXE2}
\centering
  \begin{tabular}{ll}
    \hline
    \filetext{int} & call \filetext{int} of every t-directory of the current class and
    stores \\
    & the total cross-section in file \filetext{fort/fresult.dat}\\

    \filetext{gevent} & generates and stores parton-level $\eta_b$ events of current class \\
     &  in file \filetext{fort/pdata1.dat} \\

    \filetext{fdcpythia} & generates hadronic events of the current class using \\
    &cross-section of each process as weight and stores \\
    & in file \filetext{fort/pyresult.dat} \\

    \filetext{parameter.input} & input-file in which users can specify intimal
    parameters\\ &  for the current run level \\

    \filetext{pararefresh1} & pass parameter RESET in level 2 to level 3 \\
    & and overwrite its value in level 3\\

    \filetext{pararefresh2} & pass parameter RUNNINGLEVEL in level 2 to level 3 \\
    & and overwrite its value in level 3\\

    \filetext{pararefresh3} & pass parameter PARTONEVNTNUM in level 2 to level 3 \\
    & and overwrite its value in level 3\\
    \hline

  \end{tabular}
\end{table}

The program \filetext{fdcpythia} uses cross-section as weight to
determine the number of events for each process in the current
class. And then each process calls its own \filetext{fdcpythia} for
the event generation.After executed \filetext{fdcpythia},user should
go back to \filetext{ggetab\textit{n}/f\textit{n}} to execute
\filetext{mvdir} to generate the hbook document for the annihilation
or fragmentation process

The content of the configuration-file \filetext{parameter.input} is
the same as the one at level 3 except the parameter
\texttt{RUNNINGLEVEL} should be set to 2.

\subsection{Running level 1}

When EtabFDC runs at level 1, both annihilation and fragmentation
processes are enabled in the event generation. There are six
programs \filetext{int}, \filetext{gevent}, \filetext{fdcpythia},
\filetext{pararefresh1}, \filetext{pararefresh2}  and
\filetext{pararefresh3} stored in directory
\filetext{ggetab\textit{n}/fort} and function similarly as their
counterparts at level 2. After executed \filetext{fdcpythia},user
should also go back to \filetext{ggetab\textit{n}} to execute
\filetext{mvdir} to generate the hbook document.The input-file
\filetext{parameter.input} is also kept in
\filetext{ggetab\textit{n}/fort}, and its content is the same as the
one at level 2 except that the parameter \texttt{RUNNINGLEVEL}
should be set to 1.

\subsection{Interface with PYTHIA}\label{subsecINTERFACE}
\begin{figure}[!h]
\centering
\includegraphics[scale=0.5]{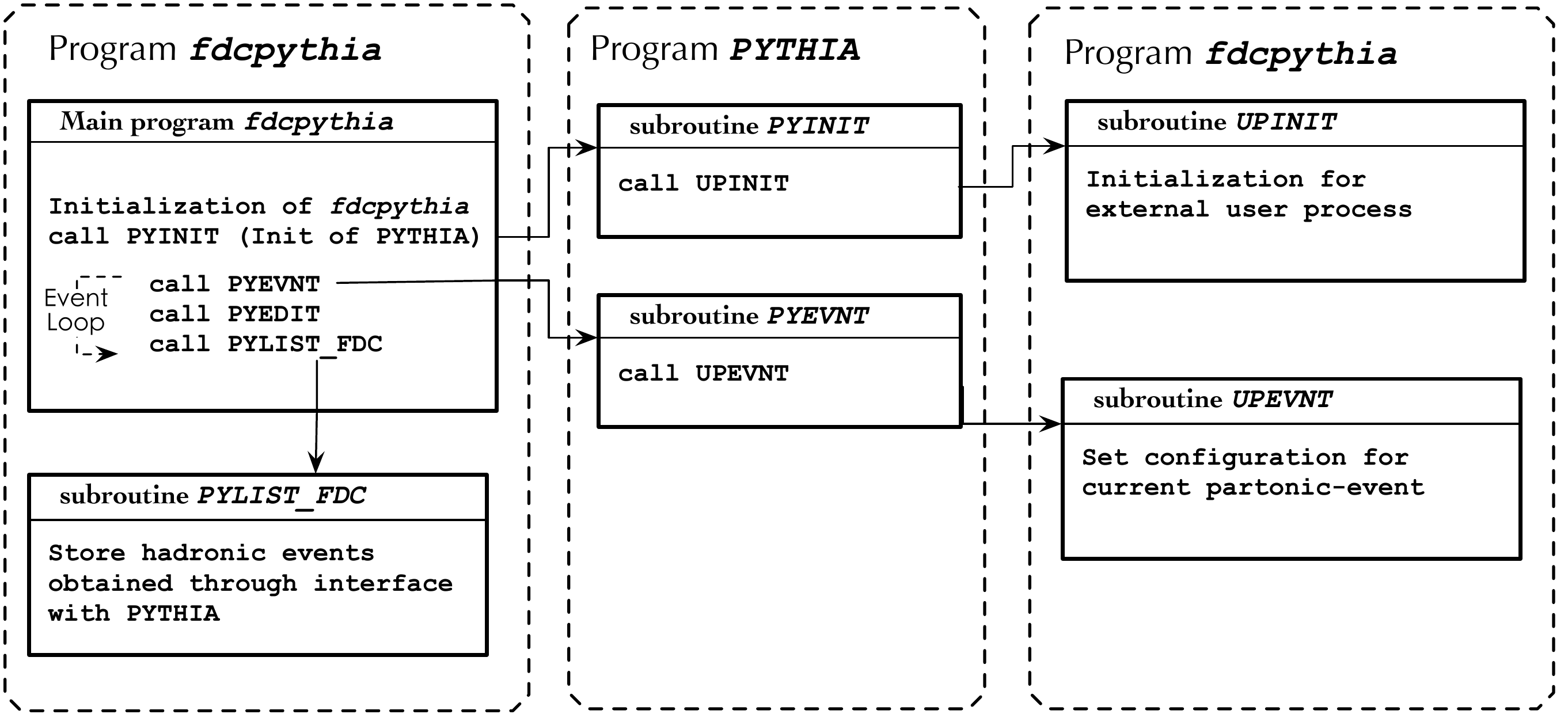}
\caption{Interface with \pythia}\label{figPYTHIA}
\end{figure}

The EtabFDC generates parton-level events which consists of
four-momenta and color flow information of all particles involved in
a process. The event data are then stored in file
\filetext{pdata1.dat} in corresponding directory
\filetext{t\textit{n}/fort} (see section \ref{secEVENTFILE} for the
format of \filetext{pdata1.dat}). In turn, like many other matrix
element generators such as \texttt{AMEGIC++} \cite{Krauss:2001iv}
and \texttt{AcerMC} \cite{Kersevan:2004yg}, EtabFDC interfaces with
\pythia for parton showering and hadronization. This interface
complies with \pythia standard and users can refer to \pythia
manual\cite{pythia}. Figure \ref{figPYTHIA} shows the general
procedures of the interface, and the source codes below is the event
loop part in program \filetext{fdcpythia}

\begin{verbatim}
C... Event loop
        DO IEV=1,NEVNT
           CALL PYEVNT()
           CALL PYEDIT(1)
           IRECS=PYLIST_FDC(IFU2,IRECNO)
           IRECNO=IRECNO+IRECS
        ENDDO
\end{verbatim}

A hadronic event is generated in a single run of the event loop.
First, the \pythia subroutine \subroutinetext{PYEVNT} calls another
subroutine \subroutinetext{UPEVNT} (which is provided by EtabFDC) to
input a parton-level event. Then \subroutinetext{PYEVNT} showers and
hadronizes the parton-level event and stores the consequent hadronic
event in common block \subroutinetext{COMMON/PYJETS/}. Next, another
\pythia subroutine \subroutinetext{PYEDIT} is called to remove from
the event record the jets or particles that have fragmented or
decayed. Finally, a function \subroutinetext{PYLIST\_FDC}, which is
a modified version of \pythia subroutine \subroutinetext{PYLIST},
writes the \subroutinetext{COMMON/PYJETS/} into file
\filetext{pyresult.hbk} in ntuple mode.

\section{Usage}\label{secHOWTOUSE}

Users can choose the running level accordingly to their own needs.
Instructions for running EtabFDC at the three running levels are
given below.

\subsection{For running level 3}                                   It will be relatively convenient to add new
physical processes in the future as well.

In this running level, EtabFDC can generate events for a specific
process. To do this, follow the steps below:
\begin{enumerate}
  \item Decide the physical process for which EtabFDC will generate events;
  \item change current directory to the sub-directory
  \filetext{fort} under it (eg. \filetext{ggetab1/f2/t1/fort});
  \item Execute program \filetext{make clean} to remove old links;
  \item Execute program \filetext{./make} to link and \filetext{./int}
  to calculate the cross-section;
  \item Edit \filetext{parameter.input} and set parameters:
  \begin{itemize}
    \item \filetext{RUNNINGLEVEL=3};
    \item \texttt{PARTONEVNTNUM} and \texttt{EVNTNUM} according to users'
    needs;
  \end{itemize}
    \item Execute program \filetext{./fdcpythia}.
\end{enumerate}
    With the steps above, \texttt{EVNTNUM} hadronic events are then stored in file
    \filetext{pyresult.hbk}.

\subsection{For running level 2}
In this running level, EtabFDC can generate events for either
annihilation or fragmentation class of processes. To do this, follow
the steps below:
\begin{enumerate}
\item Decide which class of process EtabFDC will generate events for and change current directory to the f-directory
   (eg. \filetext{ggetab2/f2});
\item Execute program \filetext{./mkclean} to remove old links;
\item Execute program \filetext{./domake} to link and \filetext{./doint}
  to calculate the cross-section;
\item Change current directory to \filetext{ggetab2/f2/fort}
\item Edit \filetext{parameter.input} in current directory:
\begin{itemize}
    \item \filetext{RUNNINGLEVEL=2};
    \item \texttt{PARTONEVNTNUM} and \texttt{EVNTNUM} according to users'
    needs;
\end{itemize}
and then execute \filetext{./pararefresh1},
\filetext{./pararefresh2} and \filetext{./pararefresh3} in directory
\filetext{ggetab2/f2/fort}.
\item Execute \filetext{./fdcpythia} for hadronic events.
\item Go back to \filetext{ggetab2/f2} and execute \filetext{./mvdir} to get hbook document for the fragmentation processes.
\end{enumerate}
    With the above steps, the exact number of generated hadronic events in this
    run is given on screen and the data for these hadronic events are then
    stored in \filetext{f2.hbk} under \filetext{ggetab2/f2} directory.

\subsection{For running level 1}
In this running level, EtabFDC can generate events for all the
implemented LO $2\rightarrow2$ processes for $pp\rightarrow\eta_b
X$. To do this, follow the steps below:
\begin{enumerate}
\item Change the current directory to the sub-directory
  \filetext{fort} under it (eg. \filetext{ggetab2});
\item Execute program \filetext{./mkclean} to cancel old links;
\item Execute program \filetext{./domake} to link and \filetext{./doint}
  to calculate the cross-section;
\item Change current directory to \filetext{ggetab2/fort}

\item Edit \filetext{parameter.input} in current directory:
\begin{itemize}
    \item \filetext{RUNNINGLEVEL=1};
    \item \texttt{PARTONEVNTNUM} and \texttt{EVNTNUM} according to users'
    needs;
  \end{itemize}
and then execute \filetext{./pararefresh1},
\filetext{./pararefresh2} and  \filetext{./pararefresh3} in
directory \filetext{ggetab2/fort}.

\item Execute \filetext{./fdcpythia} for hadronic events.

\item Go back to \filetext{/ggetab2} and execute \filetext{./mvdir} to get hbook document for all LO $2\rightarrow2$ processes for $pp\rightarrow \eta_b+X$.
\end{enumerate}
    With steps above, the exact number of generated hadronic events in this run
    is given on screen and the data for these hadronic events are then
    stored in \filetext{ggetab2.hbk} under \filetext{ggetab2} directory.

\section{Output files}\label{secEVENTFILE}

As described above, the EtabFDC interfaces with \pythia directly at
level 3. The partonic events information are kept in files named
\filetext{pdata1.dat} of the corresponding directories
\filetext{t\textit{n}/fort} in ASCII format, and the hadronic events
information are in file \filetext{pyresult.dat} in BINARY format in
the same directory.

\subsection{Format of \texttt{pdata1.dat}}\label{subsecPDATA1}
File \filetext{pdata1.dat} is generated by program
\filetext{gevent}. In this file, parton-level events are recorded in
ASCII format sequentially. For a $2\rightarrow n$ process, each
partonic event in the file \filetext{pdata1.dat} occupies $2+n$
lines (if not colored) or $2+n+1$ lines (if colored). The first
$2+n$ lines are four-momenta $(\vec{p},E)$ of initial and final
particles. The $(2+n+1)$th line is the coefficients of color bases,
which are used as weight when there is a color-flow multiplicity.
The color-flow assignment complies with the \pythia standard and
users can refer to \pythia manual\cite{pythia}. In this version of
EtabFDC, no subroutines are provided for accessing to the
parton-level events. The \subroutinetext{FORMAT} statements which
are used in \filetext{gevent} to output the four-momenta and the
coefficients of color-flow bases is given below

\begin{tabular}{|c|}
\hline \hspace*{3em}   \texttt{FORMAT(2X,4(F14.8,1X))}
\hspace*{3em} \\
   \texttt{FORMAT(2X,$\mathcal{N}$(F14.8,1X))}\\
\hline
\end{tabular}

where $\mathcal{N}$ is the number of color bases.

\subsection{Format of \texttt{pyresult.hbk}} \label{subsecPYRES}
File \filetext{pyresult.hbk} is created by \filetext{fdcpythia}. As
described in section \ref{subsecINTERFACE}, \pythia stores the
hadronized events in a common block as below:
\medskip

\begin{ttfamily}
\begin{tabular}{|l|}
\hline
INTEGER N,NPAD,K\\
DOUBLE PRECISION P,V\\
COMMON/PYJETS/N,NPAD,K(4000,5),P(4000,5),V(4000,5)\\
\hline
\end{tabular}
\end{ttfamily}

\bigskip Program \filetext{fdcpythia} then writes them into
\filetext{pyresult.hbk} in ntuple format. If running level 1 or
level 2,after executed \filetext{./fdcpythia},the user should go
back to \filetext{ggetab\textit{n}}(for level 1) or
\filetext{ggetab\textit{n}/f\textit{n}}(for level 2) and execute
\filetext{./mvdir} to get the hbook document for level 1 or level 2
seperately.The hbook document could be read by \filetext{paw}
directly.
\bigskip
\noindent

\section{Summary}\label{secCON}
We have presented a FORTRAN program package EtabFDC which is a
matrix element event generator for simulating the production and the
decays of $\eta_b$. It has implemented all possible leading-order
$2\rightarrow2$ processes and can generate parton-level events
through three running levels, and then interfaces with \pythia to
obtain hadronic events, which are stored in the binary files and can
be accessed by the provided FORTRAN subroutines. This package has
been tested with setting the $pp$ collisions energy at $\sqrt{s}=14$
TeV and the mass of $\eta_b$ to be $9.3$ GeV. One of the attractive
features of this package is that all its FORTRAN codes are generated
by FDC with full automation. With this feature, the errors caused by
carelessness can be systematically avoided.

\vspace{1.3cm}
\par
\par
{\bf Acknowledgments} \vspace{.2cm}

This work was supported in part by NSFC under Grant No:
10491306,10521003,10775179, in part by the Scientific Research Fund
of GUCAS with number 055101BM03 and 110200M202.

\newpage
\appendix
{\noindent \Large \bf Appendix : Usage Illustrations} \\

Users can choose the running level for their own purposes.

Following examples will be used to explain the usage of EtabFDC for
3 running levels.

\subsection{Level 3}

In this level, user can generate events for a specific process.

It is supposed to generated 1,000 hadronic events for process $q +
\bar{q} \rightarrow \ g + \eta_{b}(q=uds)$ , $\eta_b \rightarrow
J/\psi J/\psi$.

For this purpose, following steps are needed.
\begin{enumerate}
\renewcommand{\labelenumi}{\arabic{enumi})}
\item This process corresponds to directory \filetext{ggetab2/f1/t1}.

\item Change current directory to \filetext{ggetab2/f1/t1/fort}.

\item Edit \filetext{parameter.input} and set \\
 \texttt{
\hspace*{2em}EVNTNUM = 1000,\\
\hspace*{2em}RUNNINGLEVEL = 3,\\
\hspace*{2em}RESET = 1. }

\item Execute \filetext{./int} for cross-section. (optional)

\item Execute \filetext{./fdcpythia} for hadronic events.
\end{enumerate}
Remarks:\hfill
\begin{enumerate}
\renewcommand{\labelenumi}{\arabic{enumi})}
\item The step 4 is optional since the events are
generated unweighted in level 3. Therefore, if users do not want to
know the cross-section, they can omit this step.

\item When \filetext{./int} runs successfully, it will give no
message on screen. However, if the mass of final-particles are
greater than that of initial particles, which means the process is
prohibited by current energy configuration, \filetext{./int} prints
a message, ``\texttt{The center mass energy < the sum of all final
mass}". When this message appears, users should not execute
\filetext{./fdcpythia} or \filetext{./gevent}.

\item As mentioned in section \ref{secImp}, the parton-level events
are generated by \filetext{./gevent}. However, as shown in steps
above, \filetext{./gevent} is not used explicitly. That is because
\filetext{./fdcpythia} can execute \filetext{./gevent} automatically
according to parameters set in \filetext{parameter.input}. Of
course, if users are only interested in the parton-level events,
they can execute \filetext{./gevent} manually instead of
\filetext{./fdcpythia}.

\item Parton-level and hadronic events are stored in
\filetext{pdata1.dat} and \filetext{pyresult.hbk} respectively in
the directory \filetext{ggetabn/f1/t1/fort}. The access to them is
formulated in section \ref{secEVENTFILE}.
\end{enumerate}

\subsection{Level 2}

In this level, users can generates events for the annihilation or
fragmentation processes followed by which $\eta_{b}$ decays in
certain channel.

As an example, we take annihilation-etab-generating processes decay
as $\eta_b \rightarrow J/\psi\ + J/\psi$. It is assumed to generate
5,000 hadronic events. For this purpose, users need to go through
steps below.
\begin{enumerate}
\renewcommand{\labelenumi}{\arabic{enumi})}
\item The annihilation-etab-generating processes correspond to
\filetext{ggetab2/f1}, then go into this directory.

\item Execute shell script \filetext{./doint}.

\item Change current directory to \filetext{ggetab2/f1/fort}

\item Edit \filetext{parameter.input} in current directory and set \\
 \texttt{
\hspace*{2em}EVNTNUM = 5000,\\
\hspace*{2em}RUNNINGLEVEL = 2,\\
\hspace*{2em}RESET = 1. }\\
and then execute \filetext{./pararefresh1},
\filetext{./pararefresh2} and \filetext{./pararefresh3} in directory
\filetext{ggetab2/f1/fort}.

\item Execute \filetext{./fdcpythia} for hadronic events.

\item Go back to \filetext{ggetab2/f1} and execute
\filetext{./mvdir} to get the hbook document
\end{enumerate}
Remarks:\hfill
\begin{enumerate}
\renewcommand{\labelenumi}{\arabic{enumi})}
\item Users should notice that in level 2 (also in level 1), shell
scripts should be used instead of executable programs
\filetext{./int}.

\item In level 2 (also in level 1), the execution of \filetext{./doint}
is a necessary step since the generator needs the cross-section as
weight to determine the number of events for each specific process.

\item As in level 3, parton-level events are stored
in sub-directory \filetext{fort} of corresponding t-directories
under \filetext{ggetab2/f1}. The access to them has been described
in section \ref{secEVENTFILE}.
\end{enumerate}

\subsection{Level 1}

In this level, users can generate events for all processes. We take
the $\eta_b \rightarrow J/\psi \gamma$ decay channel as an example.
It is assumed that 10,000 hadronic events are generated. For this
purpose, following steps are necessary.

\begin{enumerate}
\renewcommand{\labelenumi}{\arabic{enumi})}
\item Change current directory to \filetext{ggetab1}.

\item Execute shell script \filetext{./doint}.

\item Change current directory to \filetext{ggetab1/fort}

\item Edit \filetext{parameter.input} in current directory.
Set \\
 \texttt{
\hspace*{2em}EVNTNUM = 10000,\\
\hspace*{2em}RUNNINGLEVEL = 1,\\
\hspace*{2em}RESET = 1. }\\
and then execute \filetext{./pararefresh1},
\filetext{./pararefresh2} and \filetext{./pararefresh3} in directory
\filetext{ggetab1/fort}.

\item Execute \filetext{./fdcpythia} for hadronic events.
\item Go back to \filetext{ggetab1} and execute \filetext{./mvdir} to get the hbook
document.
\end{enumerate}
Remark:\hfill

Except the directory, everything is the same as in level 2. The
parton-level event files are also kept in \filetext{fort} of
corresponding t-directories.

\end{document}